\documentstyle[11pt,newpasp,twoside,epsf]{article}
\def\sss{\scriptscriptstyle}

\def\Teff{T_{\rm eff}}
\def\sqr#1#2{{\vcenter{\vbox{\hrule height.#2pt 
  \hbox{\vrule width.#2pt height#1pt \kern#1pt 
  \vrule width.#2pt} 
  \hrule height.#2pt}}}}

\begin{document}

\title{Non-equilibrium chemistry in the atmospheres of brown dwarfs}
\author{D. Saumon}
\affil{Department of Physics and Astronomy, Vanderbilt University,
       Nashville, TN 37235}
\author{M. S. Marley}
\affil{NASA Ames Research Center, Moffett Field, CA 94035}
\author{K. Lodders}
\affil{Dept. of Earth \& Planetary Science, Washington University, St-Louis, MO 63130-4899}
\author{R. S. Freedman}
\affil{The Space Physics Research Institute, NASA Ames Research Center,
       Moffett Field, CA 94035}

\begin{abstract}
Carbon monoxide and ammonia have been detected in the spectrum of Gl 229B at abundances
that differ substantially from those obtained from chemical equilibrium.  Vertical
mixing in the atmosphere is a mechanism that can drive slowly reacting species 
out of chemical equilibrium.   We explore the effects of vertical
mixing as a function of mixing efficiency and effective temperature on the
chemical abundances in the atmospheres of brown dwarfs and on their spectra.  The
models compare favorably with the observational evidence and indicate that vertical
mixing plays an important role in brown dwarf atmospheres.
\end{abstract}

\section{Introduction}
The discovery of strong methane bands in the spectrum of Gl 229B 
indicated at once that it has a very low effective temperature ($\sim 1000\,$K)
and it was hailed as the first clearly identified brown dwarf (Nakajima et al. 1995;
Oppenheimer et al. 1995).
This follows from the chemistry of carbon in stellar atmospheres. In cool
M dwarfs, carbon is predominantly in the form of carbon monoxide (CO), while in Jovian 
planets methane (CH$_4$) is the dominant form.  The transition occurs in the brown 
dwarf regime, at $T \sim 1100\,$K for a pressure of 1 bar (Fegley \& Lodders 1996).

The presence of methane bands is so striking that it
prompted the creation of the new spectral class of T dwarfs (Kirkpatrick et al.
1999).  Gl 229B is a T6 dwarf.  
In this context, the discovery of a strong CO band in the 4.5 -- 5$\,\mu$m spectrum of Gl 229B
by Noll, Geballe \& Marley (1997) is 
quite remarkable.  Analysis of the spectrum reveals an abundance of CO that is over
3 orders of magnitude larger than expected from chemical equilibrium calculations
(Noll et al. 1997; Griffith \& Yelle 1999; Saumon et al. 2000), as anticipated by 
Fegley \& Lodders (1996).  A similar situation is
observed in the atmosphere of Jupiter where CO is detected while chemical 
equilibrium overwhelmingly favors CH$_4$ as the reservoir of carbon (Prinn \& Barshay 1977).

The excess of CO in Gl 229B can be understood if the chemistry of
carbon is driven away from equilibrium by vertical mixing on a time scale shorter than
the rate of conversion of CO into CH$_4$ (Fegley \& Lodders 1996; Noll et al 1997; 
Griffith \& Yelle 1999).

The two reservoirs of nitrogen in brown dwarfs are molecular nitrogen (N$_2$) and
ammonia (NH$_3$). Ammonia is the dominant molecule at low temperatures.  While N$_2$ is
invisible in the near infrared, Saumon et al.
(2000) detected NH$_3$ features in Gl 229B in the $K$ band and inferred its presence
in the $H$ band.  Each detection corresponds to a different level in the atmosphere.
NH$_3$ abundances determined separately for each band reveal a value in agreement
with chemical equilibrium in the $H$ band and a {\it depletion} by at least a factor of 4
in the $K$ band.  These results can be interpreted consistently with the CO observation
by invoking vertical mixing in the atmosphere on a time scale shorter than the rate of 
conversion of N$_2$ into NH$_3$.

The modeling of non-equilibrium chemistry of CO and other molecules due to
mixing in planetary atmospheres is a mature field (e.g. Barshay \& Lewis 1978; Fegley \& Prinn 1985;
Fegley \& Lodders 1994; Lodders \& Fegley 1994).  Application to brown dwarfs
have so far been limited to modeling the excess of CO in Gl 229B (Griffith \& Yelle 1999)
and a detailed study of the kinetics of CNO chemistry (Lodders \& Fegley 2002).
Here we present the first systematic 
exploration of the effects of mixing on the chemistry of carbon and nitrogen in brown 
dwarfs atmospheres.  We first summarize the main features of non-equilibrium chemistry 
of carbon and nitrogen caused by vertical mixing.
Using a simple parametrization for the rate of mixing, we then compute the resulting 
non-equilibrium abundances of CO, CH$_4$, N$_2$, NH$_3$ and H$_2$O as a function of
effective temperature and mixing efficiency.  The new abundance profiles lead to
significant changes in the emergent spectra of cool brown dwarfs.

\section{Carbon and Nitrogen Chemistry}

The relative abundances of CO and CH$_4$ can be driven out of equilibrium in the cool atmospheres
of brown dwarfs and giant planets because of the strong asymmetry in the reaction rates of the
(net) reaction
$${\rm CO} + 3 {\rm H}_2 \leftrightarrow {\rm CH}_4 + {\rm H}_2{\rm O}.\eqno(1)$$
Because of the large binding energy of CO, the reaction proceeds much more slowly to the right than
to the left. Vertical mixing dredges up hot gas that is relatively
rich in CO to the cooler part of the atmosphere, where CO is converted very slowly into CH$_4$, 
while the CH$_4$ 
that is brought down is quickly converted into CO.  The net result is an overabundance of CO,
with the CO/CH$_4$ ratio in the upper atmosphere fixed to its value at the level 
where the mixing time scale and the chemical time scale are equal.  
Nearly all of the elemental oxygen is partitioned between condensates, carbon monoxide 
and water.  It follows that an overabundance of CO implies an underabundance of H$_2$O.

The conversion of N$_2$ (favored at higher temperatures) into NH$_3$ is entirely analogous
to that of CO/CH$_4$. The net reaction is
$${\rm N}_2 + 3{\rm H}_2 \leftrightarrow 2 {\rm NH}_3.\eqno(2)$$
The large binding energy of N$_2$ leads to a slow reaction rate to the right, while the 
reaction to the left proceeds rapidly. 
Thermochemical and kinetic data indicate that all other molecular and atomic species identified in
the spectra of brown dwarfs react on very short time scales and are expected to be
in chemical equilibrium.  
In brown dwarfs, non-equilibrium chemistry induced by vertical mixing is thus limited to
the five species CO, CH$_4$, H$_2$O, NH$_3$ and N$_2$. 

The correct chemical pathway and reaction time scale for the conversion of CO into CH$_4$ 
remains somewhat uncertain (Griffth \& Yelle 1999; Lodders \& Fegley 2002).  For
the purposes of this exploratory calculation, however, this is not a critical consideration
and we adopt the chemical time scales for the net reactions (1) and (2) given in Lodders \& 
Fegley (2002).

\section{Vertical Transport}

In radiative atmospheres, vertical mixing can occur by eddy turbulence.  This process is 
analogous to diffusion and occurs over a characteristic time scale
$$\tau_{\rm\sss mix} \sim {H^2 \over K},\eqno(3)$$
where $H$ is the pressure scale height and $K$ is the coefficient of diffusion.  While 
$K$ is a free parameter in our calculation,  it ranges from $\sim 10^2$ to $10^5\,$cm$^2$/s 
in planetary stratospheres.

Atmospheres of brown dwarfs all become convective below a certain
depth and convection is a very effective mixing mechanism.  The time scale
for convective mixing is
$$\tau_{\rm\sss mix}=\tau_{\rm\sss conv} \sim {H_c \over v_c} = {H^2 \over K_c} ,\eqno(4)$$
where $H_c$ and $v_c$ are the convective mixing length and velocity, respectively.  
For convenience, we choose $H_c=H$.
An effective diffusion coefficient, $K_c$, can be defined for convective mixing 
(Eq. 4) and evaluated using the mixing length formalism.  In brown dwarf 
atmospheres, we find $K_c \sim 10^8 - 10^9\,$cm$^2$/s.  Thus,
the transition from a convective zone to a radiative zone in the atmosphere is accompanied
by a sudden decrease in the efficiency of mixing.

\begin{figure}[t]
\plotfiddle{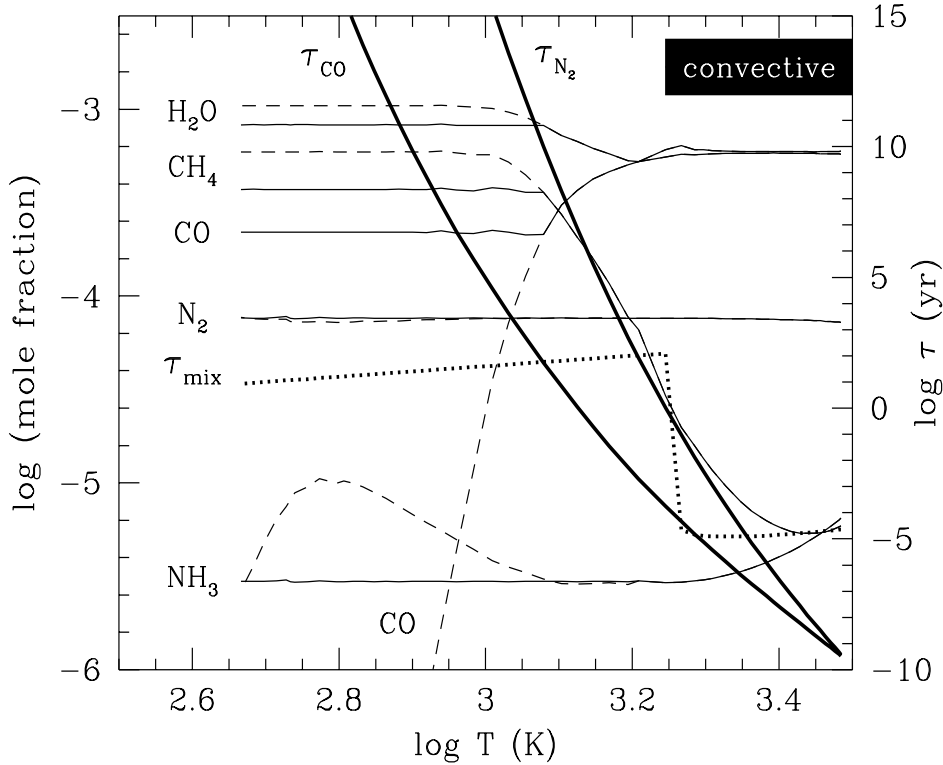}{9.9cm}{0}{67}{72}{-200}{-145}
\caption{Chemistry inside a $\Teff=1200\,$K, $g=10^5\,$cm/s$^2$ cloudless atmosphere.
         Dashed lines show the equilibrium abundances of molecules of interest and
         solid lines show the abundances that result from non-equilibrium chemistry
         driven by vertical transport.  In the radiative zone, the eddy diffusion
         coefficient is set to $K=100\,$cm$^2$/s. The extent of the convection zone is
         indicated in the upper right corner.} 
\end{figure}

\section{Non-Equilibrium Chemistry in Brown Dwarf Atmospheres}

We model the effects of vertical transport on the chemistry of carbon and nitrogen
in {\it cloudless} atmospheres with $700 \le \Teff \le 2000\,$K, $g=10^5\,$cm/s$^2$, and solar 
abundances.  The eddy diffusion coefficient in the
radiative zone is varied from $K=10^2$ to $10^{10}\,$cm$^2$/s while in the convective
zone the mixing time scale is calculated from the mixing length formalism.

An example of the effect of vertical mixing on the chemical abundance profiles in 
a model atmosphere is shown in Figure 1.  Depth in the atmosphere is indicated
by the local temperature.  The equilibrium abundances, shown with dashed lines, 
clearly show the transition in carbon chemistry from CO deep in the atmosphere 
to CH$_4$ near the surface.  The oxygen freed by the conversion of CO into CH$_4$ toward 
the top of the atmosphere is used to form H$_2$O.  The H$_2$O abundance thus increases 
in concert with that of CH$_4$.  In this relatively hot model, nearly all nitrogen is 
in the form of N$_2$ and the NH$_3$ abundance remains very low throughout the atmosphere.

\begin{figure}[t]
\plotfiddle{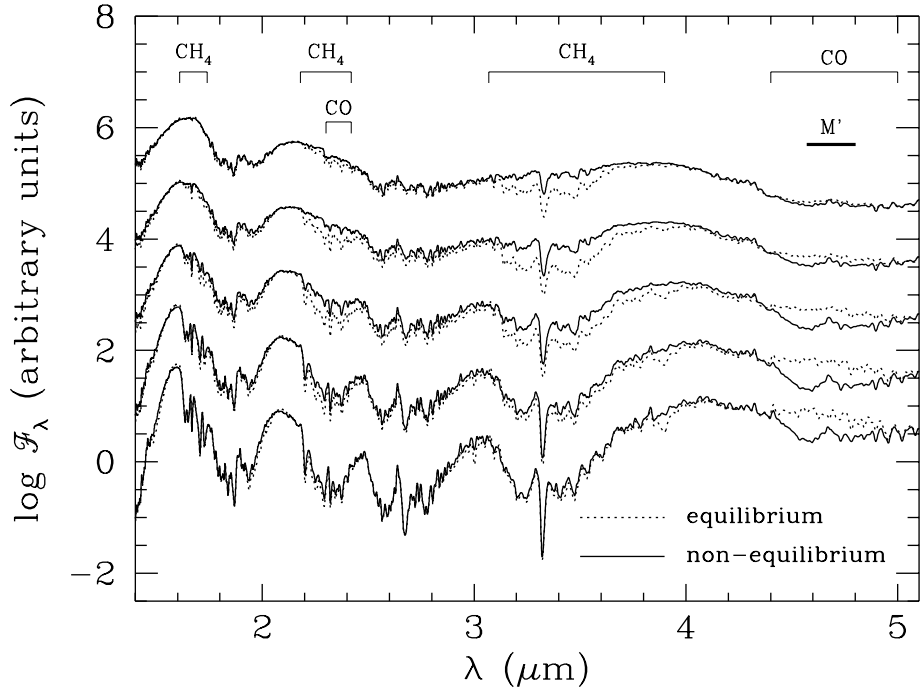}{9.0cm}{-90}{60}{60}{-250}{325}
\caption{A comparison of spectra computed with abundances from chemical equilibrium 
         and with the non-equilibrium abundances resulting from vertical transport.
         All models have $g=10^5\,$cm/s$^2$ and $K=10^4\,$cm$^2$/s. $\Teff$ decreases
         from 1600$\,$K (top) to 800$\,$K (bottom) in steps of 200$\,$K.  The spectra are
         plotted at a resolution of $R=200$ and are offset vertically for clarity.
         The main bands of CO and CH$_4$  and the bandpass of the $M^\prime$ filter
         are indicated.}
\end{figure}

The chemical time scales  for the conversion of CO and N$_2$ are shown by the two heavy 
solid lines.  These vary by more than 25 orders of magnitude throughout the atmosphere!
Mixing occurs rapidly in the convection zone but is considerably less efficient in the 
radiative zone for our choice of $K=100\,$cm$^2$/s.  In regions where 
$\tau_{\rm\sss chem} < \tau_{\rm\sss mix}$, chemical equilibrium prevails. Where 
$\tau_{\rm\sss chem} > \tau_{\rm\sss mix}$, however, the mole fractions are determined by their
values at the crossing point.  As can be seen in Fig. 1, there can be more 
than one level
where $\tau_{\rm\sss chem} = \tau_{\rm\sss mix}$, resulting in alternating zones in and out
of chemical equilibrium.  For simplicity, we consider only the uppermost (lowest T) crossing.
It turns out that in all cases, this is an excellent approximation to the more complex, full
solution because the species affected have either 1) very low abundances or 2) have nearly 
flat abundance profiles where the deeper crossings occur and are unaffected by mixing.

The resulting non-equilibrium abundances are shown by solid lines in Fig. 1.  The abundance of CO 
is considerably enhanced by vertical transport while both CH$_4$ and H$_2$O are reduced by 
more modest factors.  N$_2$ is ``frozen'' at a deeper level than CO.  Its 
abundance is barely affected in this model, while the NH$_3$ abundance decreases.  The net effect of
vertical transport in all models is an increase in the abundances of CO and N$_2$, and 
a decrease in the abundances of CH$_4$, H$_2$O and NH$_3$.  Shortening the mixing time
scale in the radiative zone (i.e. increasing $K$) produces larger changes in the abundances
for fixed $\Teff$ and gravity.

The effect of non-equilibrium chemistry on the spectra of brown dwarfs is shown in Figure 2.
The change in the strength of the 4.7$\,\mu$m band of CO increases gradually toward lower
$\Teff$ up to a maximum at $\Teff \sim 900\,$K.
The corresponding depletion in CH$_4$  results in 
weaker CH$_4$ bands with a maximum effect at $\Teff \sim 1400\,$K.  The 3.3$\,\mu$m band
of CH$_4$ is the most strongly affected, followed by the 2.2$\,\mu$m and 1.6$\,\mu$m bands.

The strongest features of NH$_3$ appear in the 10--11$\,\mu$m region and have not yet
been observed in any brown dwarf.  Ammonia forms only at relatively low temperatures
($\Teff \la 1200\,$K)
and the strength of the features increases monotonically with decreasing $\Teff$.  
Vertical mixing systematically reduces the abundance of NH$_3$ compared to its equilibrium 
value.  Since NH$_3$ is detectable at three different levels in the atmosphere (in the 
$N$, $K$, and $H$ bands; Saumon et al. 2000), the flat abundance profile predicted by this model
(Fig. 1) can in principle be tested observationally.

\begin{figure}[t]
\plotfiddle{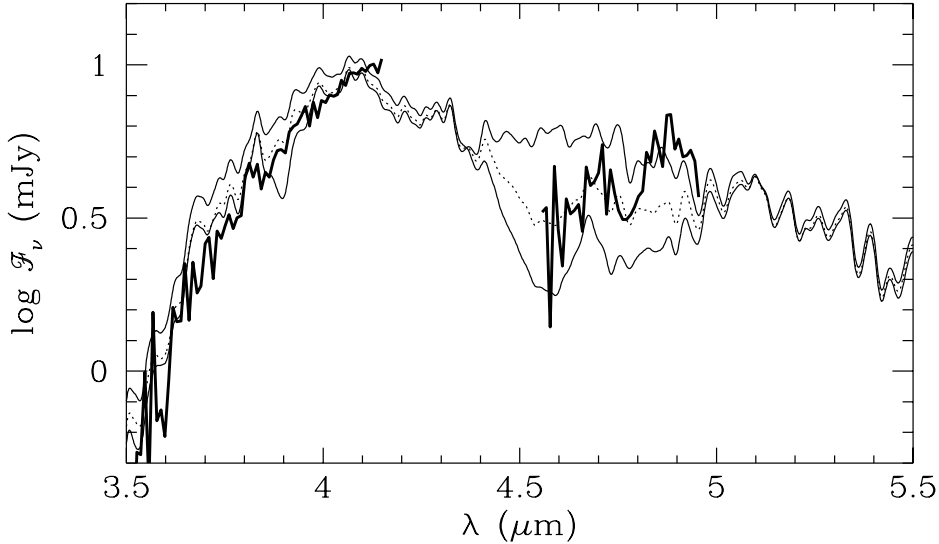}{7.3cm}{-90}{55}{55}{-235}{290}
\caption{Carbon monoxide in Gl 229B.  The spectrum of Gl 229B is
         shown as a thick solid line (Oppenheimer et al. 1998; Leggett et al. 1999).  
         The 4.5 -- 5$\,\mu$m spectrum is flux-calibrated 
         using the $M^\prime$ photometry of Golimowski et al. (2002).
         Thin curves are models with $\Teff=1000\,$K, $g=10^5\,$cm/s$^2$ and $K=0$, 
         10$^2$ and $10^4\,$cm$^2$/s, from top to bottom, respectively.  $K=0$ corresponds 
         to the  chemical equilibrium case.  The absolute flux level of the models
         has {\it not} been adjusted to fit the data.
         The models are plotted at a resolution of $R=200$.}
\end{figure}

Saumon et al. (2000) argued that the most reliable diagnostic of metallicity for T dwarfs
at present is the forest of H$_2$O lines in regions of the spectrum where water is the 
dominant absorber.  A determination
of the metallicity based on H$_2$O lines would be affected by non-equilibrium chemistry,
which can reduce its abundance by up to $\sim 0.3$ dex (increasing with $K$ and $\Teff$).  

\section{Observational Evidence}

Gl 229B is the only T dwarf with a published 4.5 -- 5$\,\mu$m spectrum (Noll et al. 1997;
Oppenheimer et al. 1998) but $M^\prime$ band photometry is now available for several other brown
dwarfs (Leggett et al. 2002; Golimowski et al. 2002).  The $M^\prime$ band is a very
good probe of the strength of the CO band (Fig. 2).  Non-equilibrium spectra
are compared with the spectrum of Gl 229B in Figure 3.  A fair agreement with the rather
noisy 4.5 -- 5$\,\mu$m spectrum is obtained for $K \sim 10^2\,$cm$^2$/s.

Absolute $M^\prime$ photometry is shown in Figure 4. 
The  equilibrium models (dashed curve) overestimate the $M^\prime$ flux at lower $\Teff$.
Non-equilibrium models can reproduce the observed fluxes quite well with 
$K \sim 10^2 - 10^4\,$cm$^2$/s.  Figure 4 suggests that increased CO absorption
resulting from vertical transport is not a peculiarity of Gl 229B but a
feature common to late L dwarfs and T dwarfs.

\section{Conclusions}

Motivated by observations in the T6 dwarf Gl 229B of a strong CO band where none was
expected and of a depletion of NH$_3$, we have explored the effects of vertical mixing
in brown dwarf atmospheres on
the most prominent molecules visible in their spectra.  Vertical
mixing, by convection or eddy turbulence, can occur fast enough to keep abundances of
CO, CH$_4$, H$_2$O and NH$_3$ from reaching their values at chemical equilibrium.
Increased efficiency of mixing leads to an enrichment of CO and a depletion of
CH$_4$, H$_2$O, and NH$_3$.  The model of vertical mixing can explain the CO and NH$_3$
observations in Gl 229B self-consistently.  Furthermore, observations of NH$_3$ in
three different bandpasses can test the model prediction of a flat abundance profile 
throughout the atmosphere (Saumon et al. 2002).  In this study, the efficiency of 
mixing is treated as a free
parameter but it can eventually be calibrated with data.

New observational evidence suggests that the overabundance of CO observed in Gl 229B
may be a common feature of late L dwarfs and of T dwarfs.  If this is indeed the case, it becomes
essential to include non-equilibrium chemistry in modeling the atmospheres and spectra of
cool brown dwarfs.  Because it affects the relative abundances of major constituents of
the atmosphere, vertical mixing further complicates the determination of the composition of 
brown dwarfs.  

Finally, vertical mixing can result in a suppression of the $M^\prime$ flux by up to
one magnitude for $\Teff$ between $\sim 700$ and 1100$\,$K (Fig. 4).  This must be
taken into account in plans to image extrasolar giant planets by capitalizing on 
their predicted surperthermal $M$ band flux (Marley et al. 1996, Burrows et al. 1997).

\begin{figure}[t]
\plotfiddle{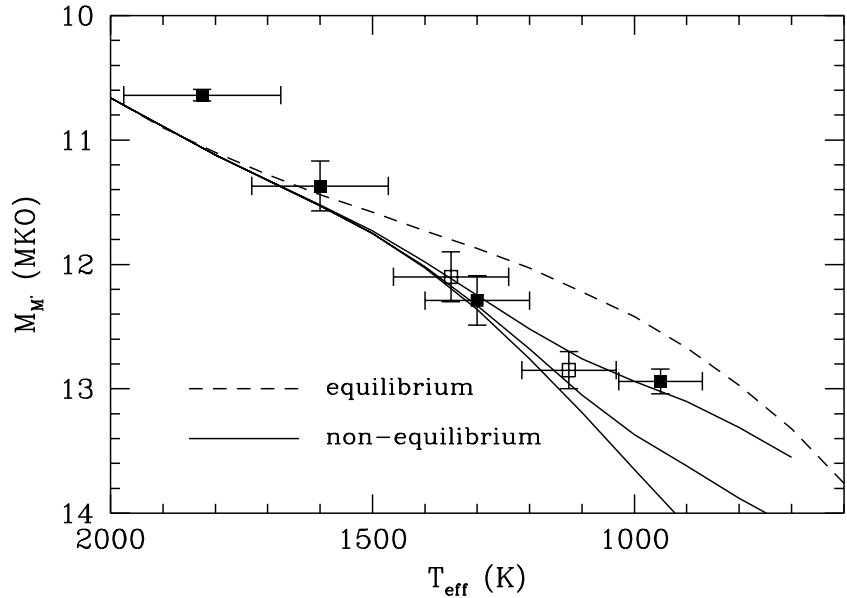}{7.8cm}{0}{67}{67}{-210}{-155}
\caption{Absolute $M^\prime$ (MKO) magnitude versus effective temperature for L and T dwarfs.
         The data are from Leggett et al. (2002) and Golimowski et al. (2002). 
         Open squares show 2MASS $0559-14$ as a single star (upper point) and as an equal pair 
         binary (lower point).  Curves show sequences
         of models with $g=10^5\,$cm/s$^2$ for different values of the eddy diffusion 
         coefficient.  From top to bottom, $K=0$, 10$^2$, $10^4$, and $10^6\,$cm$^2$/s,
         respectively.   $K=0$ corresponds to the equilibrium case.} 
\end{figure}

\acknowledgements
We thank S. K. Leggett and D. A. Golimowski for kindly providing data in 
advance of publication.  This work was supported in part by 
NSF grant AST-0086288 and NASA grant NAG5-9273 to M. S. M. and grant NAG5-4970 to
R. S. F.  Work of K. L. is
supported by NASA grant NAG5-11958 and NSF grant AST-0086487.

\end{document}